 \documentclass[prl,aps,onecolumn, showpacs]{revtex4}
\usepackage[dvips]{graphicx}
\usepackage{dcolumn}
\usepackage[english ]{babel}
\usepackage{amsmath, latexsym,  amssymb, array, graphics,
amsfonts, amsthm, amsmath,  amsfonts, array, bm, eucal}

{

\begin{document}

\newcommand{\mc}[1]{\mathcal{#1}}
\newcommand{\E}{\mc{E}}
\topmargin=-15mm
  \Large

\begin{flushright}\it\normalsize
Dedicated to the $100th$\\anniversary of the birth of \\
Nikolai Nikolaevich Bogolyubov
\end{flushright}

\title {\Large \bf  KAPITSA RESISTANCE
IN DEGENERATE QUANTUM GASES
WITH BOGOLYUBOV ENERGY EXCITATIONS
IN THE PRESENCE
OF  BOSE -- EINSTEIN CONDENSATE}

\author{\Large\bf Anatoly V. Latyshev and Alexander A. Yushkanov}

\affiliation  {Department of Mathematical Analysis and
Department of Theoretical Physics\footnote{\large N. N. Bogolyubov founded
in 1945 department of theoretical physics in Moscow State Regional
University}, Moscow State Regional
University,  105005, Moscow, Radio st., 10--A}

\begin{abstract}
The  linearized kinetic equation modelling behaviour of the degenerate
quantum bose - gas with the frequency of collisions depending
on momentum of elementary excitations is constructed.
The general case of  dependence of the  elementary excitations energy
on momentum  according to Bogolyubov formula is considered.
The analytical solution of the half--space boundary problem on
temperature  jump on border of
degenerate bose - gas in the presence of a Bose --- Einstein
condensate is received. Expression for Kapitsa resistance is received.
\\

\end{abstract}


\date{\today}
\maketitle

\section{\large 1. Introduction}

Nonequilibrium properties of the quantum gases in limited space
give rise to interest last years as well as the equilibrium  properties.
In particular, important significance  has such phenomenon,
as temperature jump \cite{1} on the border gas
-- condensed (in particular - solid)  body in the presence
 of a thermal flux normal to a surface.

Such  temperature jump is frequently called the Kapitsa
temperature jump \cite{Enz}.

The problem of temperature jump is one of
the major in the kinetic theory \cite{Cerc}.
The analytical solution of this problem
for a case of the rarefied one--atomic gas  is received in \cite {4}.

The problem of temperature jump of electronic gas in metal
was considered in our works \cite {5,6}.
In these works the analytical solution of this problem of
the temperature jump, caused by heat flux to a surface is received.

The behavior of quantum gases has aroused heigtened interest in
recent years. In particular, this is related to the development of
experimental procedures for producing and studying quantum gases at
extremely low temperatures \cite{Pit}.
The bulk properties of quantum gases
have been studied in the majority of papers
\cite{Spon} and \cite{Pang}.

At the same time, it is
obviously important to take boundary effects on the properties of such
systems into account. We mention a paper where the thermodynamic
equilibrium properties of quantum gases in a half--space were
considered \cite{Sam}.

We note that up to now, the Kapitsa jump has been calculated
in the regime where only phonon scattering at the boundary
between two media was taken into account and phonon scattering
in the bulk was neglected \cite{H}.

In the work \cite {12} the problem of temperature jump
in the quantum
Fermi -- gas was considered. The analytical solution under arbitrary
degeneration degree of the gas
has been received.
In the work \cite {13} the similar problem was considered for
Bose -- gas. But gas was assumed to be nondegenerate, i.e. in
the absence of Bose --- Einstein condensate  \cite {Lan}.

The present work is devoted to the problem of temperature jump in
degenerate Bose -- gas analysis. Presence of Bose --- Einstein
condensate leads
to essential improvement of the problem statement, and its method
of solution as well.

Thus for the description of kinetic processes near to a surface
 the kinetic equation  with model collision integral will be used.
We will assume, that boundary conditions on the
surface have specular--diffusive character.

In the present work we consider the kinetic equation, in
which Bogolyubov general dependence on  momentum for
elementary excitations energy of bose - gas in volume is taken into account.

Character of scattering of the elementary excitations on the
surface is considered thus by
introduction of phenomenological factor of reflectivity of
scattering on the surface. Thus, the given approach is
additional to \cite {H}.

In our work \cite {14} the presence fonon
component in Bogolyubov formula was neglected,
and in work \cite {15}, on the contrary,
was considered, that phonon component prevails in the elementary
excitations of bose - gas. In the given work the general case
of dependence of energy of elementary excitations
on an momentum is considered.

\section{\large 2. Derivation of the kinetic equation}

To describe the gas behavior, we use a kinetic equation with
a model collision integral analogous to that used to describe a
classical gas. We take the quantum character of the Bose gas and
the presence of the Bose --- Einstein condensate into account.

For a rarefied Bose gas, the evolution of the gas particle
distribution function $f$  can be described by the
kinetic equation \cite{Lan}
$$
\dfrac{\partial f}{\partial t} +
\dfrac{\partial \E}{\partial \mathbf{p}}
\nabla f=I[f],
\eqno{(1)}
$$
where $\E$ is the kinetic energy of gas particles, $\mathbf{p}$ is the gas
particle momentum, and $I[f]$ is the collision integral.

In the case of the kinetic description of a degenerate Bose gas,
we must take into account that the properties of the Bose --- Einstein
condensate can change as functions of the space and time coordinates,
i.e., we must consider a two--liquid model (more precisely, a
"two-fluid"\,
model because we consider a gas rather than a liquid).
We let $\rho_c=\rho_c({\mathbf{r}},t)$ and
${\mathbf{u}_c}={\mathbf{u}_c }({\mathbf{r}},t)$, thus we
denote the density and velocity of the Bose --- Einstein condensate.

We then can write the expressions \cite{H}
$$
{\mathbf{j}}=\rho_c {\mathbf{u}_c},\qquad {\mathbf{Q}}=
\dfrac{\rho_c u_c^2}{2}{\mathbf{u}_c},
$$
$$
\Pi_{ik}=\rho_c u_{ci}u_{ck}.
$$
for the densities $\mathbf{j}$, ${\mathbf{Q}}$, and $\Pi_{ik}$
of the mass, energy,
and momentum fluxes of the Bose --- Einstein condensate
(under the assumption that the chemical potential is zero) respectively.

The conservation laws for the number of particles,
energy, and momentum require that the relations
$$
\nabla {\mathbf{j}}=
-\int I[f]d\Omega_B,
$$
$$
\nabla {\mathbf{} Q}=
-\int \E({\bf p}) I[f]d\Omega_B,
$$
$$
\nabla {\Pi}=
-\int {\mathbf{p}}I[f] d\Omega_B
$$
are satisfied in the stationary case. Here,
$$
d\Omega_B=\dfrac{(2s+1)d^3p}{(2\pi \hbar)^3},
$$
$s$ is the particle spin, $\E({\bf p})$ is the energy, $\hbar$
is the Planck constant
and $I[f]$ is the collision integral in Eq. (1).

In what follows, we are interested in the case of motion with
small velocities (compared with thermal velocities). We note
that for the Bose --- Einstein condensate, the quantities ${\mathbf{Q}}$
and $\Pi_{ik}$
depend on the velocity nonlinearly (they are proportional to the
third and second powers of the velocity). Therefore, in the
linear
approximation by the velocity ${\mathbf{u}_c}$, the
energy and momentum conservation laws can be written as
$$
\int \E({\bf p})I[f]d\Omega_B=0,
$$
$$
\int {\mathbf{p}}I[f]d\Omega_B=0.
$$

According to the Bogolyubov theory, the following relation
for the excitation energy $\E(p)$ is true
for a weakly interacting Bose gas \cite{Lan}
$$
\E(p)=
\left[u_0^2p^2+\left(\dfrac{p^2}{2m}\right)^2\right]^{1/2},
\eqno{(2)}
$$
where
$$
u_0=\left(\dfrac{4\pi  \hbar^2an}{m^2}\right)^{1/2},
$$
$a$ is the scattering
length for gas particles, $n$ is the concentration,
$m$ is the mass, and $\mathbf{p}$ is the momentum of elementary
excitations, $u_0$ is the sound velocity. The parameter $a$
characterizes the interaction
force of gas molecules and
can be assumed to be small for a weakly interacting gas.

In our previous paper \cite{14}, we considered the case where
the relation
$$
u_0^2\ll \dfrac{kT}{m}
$$
is satisfied for sufficiently small $a$,
where $k$ is the Boltzmann constant and $T$ is the temperature.
In that case, the first term in the brackets in (2) can be
neglected. The expression for the energy $\E(p)$ takes the same
form as in the case of noninteracting molecules:
$$
\E(p)=\dfrac{p^2}{2m}.
$$
In this case
$$\dfrac{d \E({p})}{d\mathbf{p}}=\mathbf{v}.
$$

In our work \cite{15}, we considered the case, where
in expression (2) phonons component is prevails, i.e.,
when $T\ll \dfrac{mu_0^2}{k}$. In this case according to (2)
is received, that
$$
\E (p) =u_0 | \mathbf {p} |. % - \mathbf {p_0} |.
$$

Hence,
$$
\dfrac {d\E ({p})} {d\mathbf {p}} = u_0\dfrac {\mathbf {p}} {p}.
$$

Now we will consider the general case, when neither the first,
nor the second component in elementary excitations
can be neglected. In this case
$$
\dfrac {d \E ({p})} {d\mathbf {p}} = \alpha (p) \mathbf {p},
$$
where
$$
\alpha (p) = \dfrac {u_0^2+p^2 / (2m^2)} {\E (p)}.
$$

When considering the kinetic equation (1) under gas particles
it is necessary to understand elementary excitations of the bose - gas
with the spectrum energy (2). Character of the elementary excitations
is shown in properties of integral of collisions.
We take in the equation (1) $\tau $--approximation
as integral of its collisions.
Then the character of elementary excitations
will be shown in dependence of frequency of collisions on the
momentum of excitations \cite {Lan}, \cite {1}
$$
\dfrac{  \partial f}{ \partial t}+ \alpha(p)\mathbf{p}
\dfrac{\partial f}{\partial \mathbf{r}}=
\nu(\mathbf{p}-\mathbf{p_0})(f^*_B-f).
\eqno{(3)}
$$

Here,  $f$  is the distribution function,
$$
\nu(\mathbf{p}-\mathbf{p}_0)=\nu_0|{\bf p}- {\bf p}_0|^\gamma
$$
is the dependence of the
collision frequency on the excitation momentum,
and $\gamma$ is a constant. In the case where the phonon
component dominates in the elementary excitations,
$\gamma \geqslant 3$ \cite{Lan}. We have $\mathbf{p}_0=m{\bf v}_0$,
where  $\mathbf{v}_0$ is the velocity
of the normal component of the Bose gas, $f^*_B$ is the
equilibrium function of the Bose --- Einstein distribution
$$
f^*_B= \left[ \exp\left(\dfrac{\E(\mathbf{p}-\mathbf{p_*)}}{kT_*}
\right)-1\right]^{-1},
$$
and $\nu_0$ is a model parameter $\nu_0=\nu_1/(mkT_s)^{1/2}$, having the
meaning of the inverse mean free path $l$, $\nu_0 \sim 1/l$,
  $T_s$ is the temperature of gas in
certain point at the surface, $k$ is the Boltzmann constant,
$$
\E(\mathbf{p}-\mathbf{p}_*)=
\Big[u_0^2(\mathbf{p}-\mathbf{p}_*)^2+
\Big(\dfrac{(\mathbf{p}-\mathbf{p}_*)^2}{2m}\Big)^2\Big]^{1/2}.
$$

The parameters in $f_B^*$, namely, $T_*=T_*(\mathbf{r},t)$ and
$\mathbf{p}_*=\mathbf{p}_*(\mathbf{r},t)$,
can be determined from the
requirement that the energy and momentum conservation laws
$$
\int \nu(\mathbf{p}-\mathbf{p_0})\mathbf{p}\Big[f-f_B^*\Big]d^3p=0,
\eqno{(4a)}
$$
$$
\int \nu(\mathbf{p}-\mathbf{p_0})\E({\mathbf{p}})\Big[f-f_B^*\Big]d^3p=0
\eqno{(4b)}
$$
are applicable.

These parametres we will call an effective temperature
and an momentum respectively.

The conservation law for the number
of particles is inapplicable here because of the transition
of a fraction of particles to the Bose --- Einstein condensate.

We now assume that the gas velocity is much less than
the mean thermal velocity and the typical temperature
variations along the mean free path $l$ are small compared
with the gas temperature.
Under these assumptions, the problem can be linearized.

Let's start the linearization of equation (3).

Let's begin with linearization of the effective temperature:
$$
T_*=T_s+\delta T_*=T_s\Big(1+\dfrac{\delta T_*}{T_s}\Big).
$$

The function of Bose --- Einstein distribution is function
of the momentum $\mathbf{p}$ and parameters $\mathbf {p_*}$
and $\delta T_*/T_s$.
Its linearization we will implement by two last parameters:
$$
f_B^*(\mathbf{p},\mathbf{p_*}, \dfrac{\delta T_*}{T_s})=
f_B^*(\mathbf{p},0,0)+\dfrac{\partial f_B^*}{\partial \mathbf{p_*}}
(\mathbf{p},0,0)\mathbf{p_*}+\dfrac{\partial f_B^*}
{\partial (\frac{\delta T_*}{T_s})}(\mathbf{p},0,0)\dfrac{\delta T_*}
{T_s}.
$$

As a result of such linearization we obtain:
$$
f_B^*(\mathbf{p},\mathbf{p_*}, \dfrac{\delta T_*}{T_s})=
f_B(p)+g(p)\alpha(p)\dfrac{\mathbf{p}\mathbf{p_*}}{kT_s}+
g(p)\dfrac{\E(p)}{kT_s}\dfrac{\delta T_*}{T_s}.
\eqno{(5)}
$$

Here
$$
f_B(p)\equiv f_B^*(\mathbf{p},0,0)=\dfrac{1}{\exp\Big(\dfrac{
\E(p)}{kT_s}\Big)-1},
$$
$$
g(p)=\dfrac{\exp\Big(\dfrac{\E(p)}{kT_s}\Big)}
{\Big[\exp\Big(\dfrac{\E(p)}{kT_s}\Big)-1\Big]^2}.
$$

The linearization of the distribution function according to (5)
we will carry out as follows:
$$
f(\mathbf{r},\mathbf{p},t)=f_B(p)+g(p)h(\mathbf{r},\mathbf{p},t).
\eqno{(6)}
$$

From (5) and (6) we find, that
$$
f_B-f=g(p)\Big[\alpha(p)\dfrac{\mathbf{p}\mathbf{p_*}}{kT_s}+
\dfrac{\E(p)}{kT_s}\dfrac{\delta T_*}{T_s}\Big].
$$

Let's return to the equation (3). We will realize linearization
of this equation according to (5).

We will notice, that in linear approximation the quantity
{$\nu(\mathbf {p}-\mathbf{p_0})$}
in the equation (3) it is possible to replace by
{$\nu(p)=\nu_0p^\gamma$}.

Then the equation (3) as a result of the linearization
has the following form
$$
\dfrac{\partial h}{\partial t}+\alpha(p)\mathbf{p}\dfrac{\partial h}
{\partial \mathbf{r}}=\nu_0 p^\gamma
\Big[\alpha(p)\dfrac{\mathbf{p}\mathbf{p_*}}
{kT_s}+\dfrac{\E(p)}{kT_s}\dfrac{\delta T_*}{T_s}\Big].
\eqno{(7)}
$$

In the equation (7) we will introduce a dimensionless momentum (velocity)
$$
\mathbf{C}=\dfrac{\mathbf{v}}{v_T}=\dfrac{\mathbf{p}_T}{p_T},
$$
where $v_T=\sqrt{kT_s/m}$ is the thermal velocity of gas
particles, $\mathbf{p}_T$ is their momentum.

Then
$$
\E(p)=kT_s\E(C),
$$
where
$\E(C)=\sqrt{w_0^2C^2+C^4/4}$, and $w_0=u_0/v_T$ is
the dimensionless sound velocity.

Besides that, we will notice, that
$\alpha(p)=\alpha(C)/m$,  where
$$
\alpha(C)=\dfrac{w_0^2+C^2/2}{\sqrt{w_0^2C^2+(C^2/2)^2}}.
$$

We will introduce dimensionless time and coordinate
$\tau=\nu_0t$
and $\mathbf{r}_1=\nu_0\sqrt{\dfrac{m}{kT_s}}\mathbf{r}$.

Now it is clear, that the equation (3) (in dimensionless variables)
has the following form:
$$
\dfrac{\partial h}{\partial \tau}+
\alpha(C)\mathbf{C}\dfrac{\partial h}{\partial
\mathbf{r}_1}=$$$$=
C^\gamma\Big[\alpha(C)\mathbf{C}\mathbf{C_*}(\mathbf{r}_1,\tau)+
\E(C)\dfrac{\delta T_*}
{T_s}(\mathbf{r}_1,\tau)-h(\mathbf{r}_1,\mathbf{C},\tau)\Big].
\eqno{(8)}
$$

In laws of preservation (4) we will also implement such linearization.
As a result we get the
laws of preservation of the momentum and energy in the following form:
$$
\int p^\gamma \mathbf{p}\Big[\alpha(p)\dfrac{\mathbf{p}\mathbf{p_*}}{kT_s}+
\dfrac{\E(p)}{kT_s}\dfrac{\delta T_*}{T_s}-h(\mathbf{r}_1,
\mathbf{p},\tau)\Big]g(p)\,d^3p=0,
$$
$$
\int p^\gamma \E(p)
\Big[\alpha(p)\dfrac{\mathbf{p}\mathbf{p_*}}{kT_s}+
\dfrac{\E(p)}{kT_s}\dfrac{\delta T_*}{T_s}-h(\mathbf{r}_1,
\mathbf{p},\tau)\Big]g(p)\,d^3p=0,
$$

In these equalities we will pass to integration on a dimensionless
momentum.
We obtain:
$$
\int C^\gamma \mathbf{C}\Big[\alpha(C)\mathbf{C}\mathbf{C_*}+
\E(C)\dfrac{\delta T_*}{T_s}-h(\mathbf{r}_1,
\mathbf{C},\tau)\Big]g(C)\,d^3C=0,
$$
$$
\int C^\gamma \E(C)\Big[\alpha(C)\mathbf{C}\mathbf{C_*}+
\E(C)\dfrac{\delta T_*}{T_s}-h(\mathbf{r}_1,
\mathbf{C},\tau)\Big]g(C)\,d^3C=0,
$$
where
$$
g(C)=\dfrac{e^{\E(C)}}{\big(e^{\E(C)}-1\big)^2}.
$$

From these laws of conservation of energy and momentum it is found:
$$
\mathbf{C}_*(\mathbf{r}_1,\tau)=\dfrac{\displaystyle
\int C^\gamma \mathbf{C}h(\mathbf{r_1},\mathbf{C},\tau)g(C)\,d^3C}
{\displaystyle
\int C^\gamma C_x^2\alpha(C)g(C)\,d^3C}
$$
and
$$
\dfrac{\delta T_*}{T_s}(\mathbf{r}_1,\tau)=
\dfrac{\displaystyle
\int C^\gamma \E(C)h(\mathbf{r_1},\mathbf{C},\tau)g(C)\,d^3C}
{\displaystyle
\int C^\gamma \E^2(C)g(C)\,d^3C}.
$$

Let's calculate the integrals standing in denominators of the two last
equalities. We have:
$$
\int C^\gamma C_x^2\alpha(C)g(C)\,d^3C=\dfrac{4\pi}{3}g_1.
$$
and
$$
\int C^\gamma \E^2(C)g(C)\,d^3C=
4\pi g_2,
$$

Here
$$
g_1=\int\limits_{0}^{\infty}C^{\gamma+4}\alpha(C)g(C)\,dC,
\qquad
g_2=\int\limits_{0}^{\infty}C^{\gamma+2}\E^2(C)g(C)\,dC.
$$

Thus, equation parameters are  equal definitively
$$
\mathbf{C}_*(\mathbf{r}_1,\tau)=
\dfrac{3}{4\pi g_1}
\int C^\gamma \mathbf{C}h(\mathbf{r_1},\mathbf{C},\tau)g(C)\,d^3C,
\eqno{(9)}
$$
$$
\dfrac{\delta T_*}{T_s}(\mathbf{r}_1,\tau)=
\dfrac{1}{4\pi g_2}
\int C^\gamma
\E(C)h(\mathbf{r_1},\mathbf{C},\tau)g(C)\,d^3C.
\eqno{(10)}
$$

By means of equalities (9) and (10) we will present
the equation (8) in the standard form
for the transport theory:
$$
\dfrac{\partial h}{\partial \tau}+
\alpha(C)\mathbf{C}\dfrac{\partial h}{\partial \mathbf{r}_1}+
C^\gamma h(\mathbf{r}_1,\mathbf{C},\tau)=$$$$=
\dfrac{C^\gamma}{4\pi}\int K(\mathbf{C},\mathbf{C'})
h(\mathbf{r_1},\mathbf{C'},\tau){C'}^\gamma g(C')\,d^3C'.
\eqno{(11)}
$$

Here $K(\mathbf{C},\mathbf{C'})$ is the kernel of equation,
$$
K(\mathbf{C},\mathbf{C'})=\dfrac{3\alpha(C)\mathbf{C}\mathbf{C'}}
{g_1}+\dfrac{\E(C)\E(C')}{g_2}.
$$

\section{\large 3. Problem statement}

In the problem under consideration, a degenerate Bose
gas occupies the half--space  $x>0$ above a
planar surface where the heat exchange between
the condensed phase and the gas occurs. Therefore,
the function $h(\tau,\mathbf{r}_1,\mathbf{C})$ can be regarded as
$$
h(\tau,\mathbf{r}_1,\mathbf{C})=h(x,\mu,C).
$$
in what follows.

Here $\mu$ is cosine of a corner between a direction of a vector of
velocity $\mathbf{C}$ and an axis $x$ in spherical system of
velocities of Bose -- particles,
$C_x =\mu C$.
The equation (11) for function $h$ will be written in a form:
$$
\mu\dfrac{\alpha(C)}{C^{\gamma-1}}\dfrac{\partial h}{\partial x}+
h(x,\mu,C)=\dfrac{3\alpha(C)C\mu}{2g_1}W_1(x)+
\dfrac{\E(C)}{2g_2}W_2(x).
\eqno{(12)}
$$

In this equation
$$
W_1(x)=\int\limits_{-1}^{1}
\int\limits_{0}^{\infty}{C'}^{\gamma+3}\mu'h(x,\mu',C')g(C')d\mu'dC'
$$
and
$$
W_2(x)=\int\limits_{-1}^{1}
\int\limits_{0}^{\infty}\E(C')h(x,\mu',C'){C'}^{\gamma+2}g(C')d\mu'dC'.
$$

The problem consists in finding of quantity of the relative
temperature jump $\varepsilon_T =\Delta T/T_s, \; \Delta
T=T_s-T$,
as function $Q_x$ which is the quantity of a heat flow
projection to an axis $x$.
Considering the linear character of the problem,
it is possible to write down:
$$
\varepsilon_T=R Q_x.
$$

The dimensionless  factor $R$ of the temperature jump is called
Kapitsa resistance.

It is obvious that equation (12) has the particular solutions:
$$
h_1(x,\mu,C)=\alpha(C)C\mu
$$
and
$$
h_2(x,\mu,C)=\E(C),
$$
and the Chapmen --- Enskog distribution function is
$$
h_{as}(x,\mu,C)=B^+\alpha(C)C\mu -\varepsilon_T \E(C),
$$
where the quantity  $B^+$ is proportional to the heat flux $Q_x$.

Assuming that the reflection of the elementary
excitations from the wall is specular--diffuse,
we now formulate the boundary conditions
$$
h(0,\mu,C)=qh(0,-\mu,C),\quad 0<\mu<1,
\eqno{(13)}
$$
and
$$
h(x,\mu,C)=B^+\alpha(C)C\mu -\varepsilon_T \E(C)+o(1),\;
x\to +\infty,\;-1<\mu<0,
\eqno{(14)}
$$
where  $q$ is the specular reflection coefficient.

The problem is to solve Eq. (12) with boundary conditions (13) and (14).
Finding the value of the temperature jump $\varepsilon_T$
is of special interest.

\section{\large 4. Reduction to the integral equation}

We continue the function $h(x,\mu,C)$ to
the half--space $x<0$ symmetrically:
$$
h(x,\mu,C)=h(-x,-\mu,C), \qquad x<0.
$$

For $x<0$, we then have the following Chapmen --- Enskog
distribution
$$
h_{as}(x,\mu,C)=B^-\mu-\varepsilon_TC,
$$
with $B^+=-B^-$.

Now we will extract  the Chapmen --- Enskog
distribution of the function $h(x,\mu,C)$, considering at $ \pm x> 0$:
$$
h(x,\mu,C)=B^{\pm}\alpha(C)C\mu-\varepsilon_T\E(C)+h_c(x,\mu,C).
$$

For the function $h_c(x,\mu,C)$, we formulate the boundary conditions
for the lower and upper half--spaces:
$$
h_c(+0,\mu,C)=
$$
$$
=-(1+q)B^+\alpha(C)C\mu+(1-q)\varepsilon_T\E(C)+q
h_c(+0,-\mu,C),
$$
where $0<\mu<1$, and
$$
h_c(-0,\mu,C)=
$$
$$
=-(1+q)B^-\alpha(C)C\mu+(1-q)\varepsilon_T\E(C)+q
h_c(-0,-\mu,C),
$$
where  $-1<\mu<0$, and
$$
h_c(+\infty,\mu,C)=0, \qquad h_c(-\infty,\mu,C)=0.
$$

We include these boundary conditions in the kinetic equation.
We obtain the equation
$$
\mu\dfrac{\partial h_c}{\partial x}+\dfrac{C^{\gamma-1}}{
\alpha(C)}h_c(x,\mu,C)=
$$$$=
\dfrac{C^{\gamma-1}}{\alpha(C)}\bigg\{\dfrac{3\alpha(C)C\mu}{2g_1}
W_1(x)+\dfrac{\E(C)}{2g_2}W_2(x)+
$$$$+
|\mu|\Big[-(1+q)B^+\alpha(C)C|\mu|+(1-q)\varepsilon_T
\E(C)-$$$$-
(1-q)h_c(\mp 0,\mu,C)\big]\delta(x)\bigg\}.
\eqno{(15)}
$$

Here, $\delta(x)$ is the Dirac delta function.

Equation (15) actually combines two equations.
The point is that the term $h_c(-0,\mu,C)$
corresponds to positive $\mu: 0<\mu<1$
and the term $h_c(+0,\mu,C)$ corresponds to negative
$\mu: -1<\mu<0$.

We seek the solution of equations (15) in the form of Fourier
integrals:
$$
h_c(x,\mu,C)=\dfrac{1}{2\pi}\int\limits_{-\infty}^{\infty}
e^{ikx}\Phi(k,\mu,C)dk,\qquad
\delta(x)=\dfrac{1}{2\pi}\int\limits_{-\infty}^{\infty}
e^{ikx}dk.
$$
$$
W_1(x)=\dfrac{1}{2\pi}\int\limits_{-\infty}^{\infty}
e^{ikx}E_1(k)dk,\qquad
$$
$$
W_2(x)=\dfrac{1}{2\pi}\int\limits_{-\infty}^{\infty}
e^{ikx}E_2(k)dk.
$$

Let's begin with search of unknown boundary values
$h_c (\mp 0,\mu,C)$.
We will express these values in terms of spectral densities
$E_1(k)$ and $E_2(k)$. We will consider
for this purpose the equation (15)
for $x<0$ and $x>0$. In both cases component with
$\delta$--function drops out of the equation (15).

Solving Eq. (15) with $x>0$ and $\mu<0$, assuming
that the right--hand side of this equation is known,
and assuming that the boundary
conditions are satisfied far from the wall, we obtain
$$
h_c^+(x,\mu,C)=-\dfrac{1}{\mu}
\exp\Big(-\dfrac{x}{\mu}\dfrac{C^{\gamma-1}}{\alpha(C)}\Big)
\int\limits_{x}^{+\infty}
\exp\Big(\dfrac{t}{\mu}\dfrac{C^{\gamma-1}}{\alpha(C)}\Big)W(t,\mu,C)dt,
\eqno{(16)}
$$
where
$$
W(t,\mu,C)=\dfrac{C^{\gamma-1}}{\alpha(C)}\Bigg[\dfrac{3\alpha(C)C\mu}
{2g_1}W_1(t)+
\dfrac{\E(C)}{2g_2}W_2(t)\Bigg].
$$

For $x<0,\;\mu>0$ we similarly receive:
$$
h_c^-(x,\mu,C)=\dfrac{1}{\mu}\exp\Big(-\dfrac{x}{\mu}\dfrac{C^{\gamma-1}}
{\alpha(C)}\Big)\int\limits_{-\infty}^{x}\exp\Big(\dfrac{t}{\mu}
\dfrac{C^{\gamma-1}}{\alpha(C)}\Big)W(t,\mu,C)dt.
$$

Let's underline, that in the equation (15) boundary values of the required
functions $h_c(\pm 0,mu,C) $ are boundary values of the
represented above functions $h_c^{\pm}(x,\mu,C)$ at $x\to\pm 0$ from
corresponding semi--planes.

From two last equalities for integrals Fourier follows, that
$$
W(t,\mu,C)=\dfrac{C^{\gamma-1}}{\alpha(C)}\Bigg[\dfrac{3\alpha(C)C\mu}
{2g_1}\cdot \dfrac{1}{2\pi}\int\limits_{-\infty}^{\infty}
e^{ikt}E_1(k)\,dk+$$$$+
\dfrac{\E(C)}{2g_2}\cdot\dfrac{1}{2\pi}
\int\limits_{-\infty}^{\infty}e^{ikt}E_2(k)\,dk\Bigg].
$$

After simple calculations in (16), we obtain
$$
h_c^+(x,\mu,C)=\dfrac{C^{2(\gamma-1)}}{2\pi}\int\limits_{-\infty}^{+\infty}
e^{ikx}\Bigg[ \dfrac{3\alpha(C)C\mu}{2g_1}E_1(k)dk
+$$$$+
\dfrac{\E(C)}{2g_2}E_2(k)\Bigg]
\dfrac{dk}{C^{2(\gamma-1)}+k^2\mu^2\alpha^2(C)}.
$$

It is similarly possible to show, that for function
$h^-_c(x,\mu,C)$  precisely the same expression is received.
Hence, considering
evenness by $k$ of the functions $E_1(k)$ and $E_2(k)$, we receive, that
$$
h_c^{\pm}(0,\mu,C)=\dfrac{C^{\gamma}}{\pi}
\int\limits_{0}^{+\infty}
\Bigg[ \dfrac{3\alpha(C)C\mu}{2g_1}E_1(k)+
\hspace{4cm}
$$
$$
\hspace{4cm}
+\dfrac{\E(C)}{2g_2}E_2(k)\Bigg]
\dfrac{dk}{C^{2(\gamma-1)}+k^2\mu^2\alpha^2(C)}.
\eqno{(17)}
$$

So, boundary values of required function $h_c(\pm 0,\mu,C)$
from the equation (15) are boundary values of
$h_c^{\pm}(0,\mu,C)$, defined by equality (17).

By means of the relation (17) it is visible, that two equations (15)
are possible to be united in one:
$$
\dfrac{\mu \alpha(C)}{C^{\gamma-1}}\dfrac{\partial h_c}{\partial x}+
h_c(x,\mu,C)=
%$$$$=
\dfrac{3\alpha(C)C\mu}{2g_1}W_1(x)+
\dfrac{\E(C)}{2g_2}W_2(x)+
$$$$+
|\mu|\Big\{-(1+q)B^+|\mu|\alpha(C)C+(1-q)\varepsilon_T\E(C)\Big\}\delta(x)-
$$$$-
(1-q)|\mu|\delta(x)\dfrac{C^\gamma}{\pi}\int\limits_{0}^{\infty}
\Bigg[\dfrac{3\alpha(C)C\mu}{2g_1}E_1(k)+\hspace{1.5cm}
$$$$\hspace{1.5cm}+
\dfrac{\E(C)}{2g_2}E_2(k)\Bigg]\dfrac{dk}{C^{2(\gamma-1)}+
k^2\mu^2\alpha^2(C)}.
\eqno{(18)}
$$

\section{\large 5. Characteristic system of the equations}

We pass to the Fourier integrals in equation (18) and obtain the
equation
$$
\Big[C^\gamma+ik\mu \alpha(C)C\Big]\Phi(k,\mu,C)=$$$$=
\dfrac{3\alpha(C)C^{\gamma+1}\mu}{2g_1}E_1(k)+
\dfrac{\E(C)C^\gamma}{2g_2}E_2(k)-
$$
$$
-(1+q)B^+\mu^2C^\gamma+(1-q)\varepsilon_T C^{\gamma+1}|\mu|-
%$$$$-
(1-q)|\mu|\dfrac{C^{3\gamma}}{\pi}\times
$$$$ \times
\int\limits_{0}^{\infty}
\Bigg[\dfrac{3\alpha(C)C\mu}{2g_1}E_1(k)+
\dfrac{\E(C)}
{2g_2}E_2(k)\Bigg]\dfrac{dk}{C^{2\gamma}+
k^2\mu^2\alpha^2(C)C^2}.
\eqno{(19)}
$$

It is simple to find expressions for $E_1(k)$ and $E_2(k)$:
$$
E_1(k)=\int\limits_{-1}^{1}\int\limits_{0}^{\infty}
C^{\gamma+3}\mu \Phi(k,\mu,C)g(C)d\mu \,dC
$$
and
$$
E_2(k)=\int\limits_{-1}^{1}\int\limits_{0}^{\infty}
\E(C)C^{\gamma+2}\Phi(k,\mu,C)g(C)\,d\mu\,dC.
$$

We solve equation (19) for $\Phi(k,\mu,C)$ and substitute it in
the last two expressions.

We will introduce following designations:
$$
T^{r,s}_{m,n}(k)=\int\limits_{0}^{1}\int\limits_{0}^{\infty}
\dfrac{\alpha^r(C)\E^s(C)C^m\mu^n}{C^{2\gamma}+k^2\mu^2 \alpha^2(C)C^2}
d\mu\,dC, \quad r,s=0,1,2,\cdots,
$$
and
$$
J_{m,n}^{r,s}(k,k_1)=
\int\limits_{0}^{1}\int\limits_{0}^{\infty}
\dfrac{\alpha^r(C)\E^s(C) C^m \mu^n g(C)d\mu
dC}{(C^{2\gamma}+k^2\mu^2\alpha^2(C)C^2)
(C^{2\gamma}+k_1^2\mu^2\alpha^2(C)C^2)},
$$
where $r,s=0,1,2\cdots$.

Between these integrals the following relation is obvious:
$$
J^{r,s}_{m,n}(k,0)=T^{r,s}_{m-2\gamma,n}(k),\qquad
J^{r,s}_{m,n}(0,k_1)=T^{r,s}_{m-2\gamma,n}(k_1).
$$

By means of the entered designations we
receive characteristic system,
consisting of two equations:
$$
\Big[1-\dfrac{3}{g_1}T^{1,0}_{3\gamma+4,2}(k)\Big]E_1(k)+
\dfrac{ik}{g_2}T^{1,1}_{2\gamma+4}(k)E_2(k)=
$$
$$
=2(1+q)B^+ikT^{1,0}_{2\gamma+4,4}(k)-
2(1-q)\varepsilon_TikT^{1,0}_{2\gamma+5,3}(k)-
$$
$$
-3\dfrac{1-q}{g_1\pi}
\int\limits_{0}^{\infty}J^{1,0}_{5\gamma+4,3}(k,k_1)E_1(k_1)dk_1+
$$
$$
+ik\dfrac{1-q}{g_2\pi}\int\limits_{0}^{\infty}
J^{1,1}_{4\gamma+4,3}(k,k_1)E_2(k_1)dk_1,
$$
and
$$
\dfrac{3ik}{g_2}T^{2,1}_{2\gamma+4,2}(k)E_1(k)+
\Big[1-\dfrac{1}{g_2}T^{0,2}_{3\gamma+2,0}(k)\Big]E_2(k)=
$$
$$
=-2(1+q)B^+T^{0,1}_{3\gamma+2,2}(k)+2(1-q)\varepsilon_T
T^{0,1}_{3\gamma+3,1}(k)+
$$
$$
+3ik\dfrac{1-q}{g_1\pi}\int\limits_{0}^{\infty}
J^{2,1}_{4\gamma+4,3}(k,k_1)E_1(k_1)dk_1-
$$
$$
-\dfrac{1-q}{g_2\pi}\int\limits_{0}^{\infty}
J^{2,0}_{5\gamma+2,1}(k,k_1)E_2(k_1)dk_1.
$$

We introduce the dispersion matrix function
$$
\Lambda(k)=
\left [\setlength{\extrarowheight}{6pt}\extrarowheight=16pt
\begin{array}{cc}
1-\dfrac{3}{g_1}T^{1,0}_{3\gamma+4,2}(k)
&\dfrac{ik}{g_2}T^{1,1}_{2\gamma+4,2}(k)
\\
\dfrac{3ik}{g_1}T^{2,1}_{2\gamma+4,2}(k)&
1-\dfrac{1}{g_2}T^{0,2}_{3\gamma+2,0}(k) \\
  \end{array}
\right].
$$

From spectral densities $E_1(k)$ and $E_2(k)$ we form the vector column
$$
E(k)=\left[\extrarowheight=6pt\begin{array}{c}
           E_1(k) \\
           E_2(k) \\
         \end{array}
       \right],
$$
and we introduce two vector columns of arbitrary terms
$$
T_1(k)=
\left[\extrarowheight=6pt  \begin{array}{c}
    ikT^{1,0}_{2\gamma+4,4}(k) \\
    -T^{0,1}_{3\gamma+2,2}(k) \\
  \end{array}
\right], \qquad
T_2(k)=
\left[\extrarowheight=6pt
  \begin{array}{c}
    ikT^{1,0}_{2\gamma+5,3}(k) \\
    -T^{0,1}_{3\gamma+3,1}(k) \\
  \end{array}
\right].
$$

We will combine the received system of the scalar equations
in one vector equation
$$
\Lambda(k)E(k)=2(1+q)B^+T_1(k)-$$$$-
2(1-q)\varepsilon_TT_2(k)+
\dfrac{1-q}{\pi}\int\limits_{0}^{\infty}
J(k,k_1)E(k_1)dk_1.
\eqno{(20)}
$$

Here
$J(k,k_1)$ is the matrix kernel of integral equation (20),
$$
J(k,k_1)=
\left[\extrarowheight=19pt \begin{array}{cc}
    -\dfrac{3}{g_1}J^{1,0}_{5\gamma+4,3}(k,k_1) &
    \dfrac{ik}{g_2}J^{1,1}_{4\gamma+4,3}(k,k_1)
    \\ \hspace{16pt}
    \dfrac{3ik}{g_1}J^{2,1}_{4\gamma+4,3}
    (k,k_1) & -\dfrac{1}{g_2}J^{0,2}_{5\gamma+2,1}(k,k_1) \\
  \end{array}
\right].
$$

We will consider obvious equalities
$$
1-\dfrac{3}{g_1}T^{1,0}_{3\gamma+4,2}(k)=\dfrac{3k^2}{g_1}
T^{3,0}_{\gamma+6,4}(k),
$$
$$
1-\dfrac{1}{g_2}T^{0,2}_{3\gamma+2,0}(k)=\dfrac{k^2}{g_2}
T^{2,2}_{\gamma+4,2}(k),
$$
and we will present the dispersion matrix function in the following form:
$$
\Lambda(k)=
\left [\setlength{\extrarowheight}{6pt}\extrarowheight=16pt
\begin{array}{cc}
    \dfrac{k^2}{g_2}T^{2,2}_{\gamma+4,2}(k)
&\dfrac{3ik}{g_1}T^{2,1}_{2\gamma+4,2}(k)
\\
\dfrac{ik}{g_2}T^{1,1}_{2\gamma+4,2}(k)&
    \dfrac{3k^2}{g_1}T^{3,0}_{\gamma+6,4}(k) \\
  \end{array}
\right].
$$

\section{\large 6. Method of successive approximations}

We seek the solution of Eq. (20) in the form
$$
\varepsilon_T=\dfrac{1+q}{1-q}\Big[\varepsilon_0+\varepsilon_1(1-q)+
\varepsilon_2(1-q)^2+\cdots\Big],
\eqno{(21)}
$$
$$
E(k)=2(1+q)\Big[E^{(0)}(k)+E^{(1)}(k)(1-q)+E^{(2)}(k)(1-q)^2+\cdots\Big].
\eqno{(22)}
$$

We substitute these equalities in the characteristic equation.
We obtain the countable system of equations
$$
\Lambda(k)E^{(0)}(k)=-B^+T_1(k)+\varepsilon_0T_2(k),
\eqno{(23)}
$$
$$
\Lambda(k)E^{(1)}(k)=\varepsilon_1T_2(k)-\dfrac{1}{\pi}
\int\limits_{0}^{\infty} J(k,k_1)E_0(k_1)dk_1,
\eqno{(24)}
$$
$$
\Lambda(k)E^{(2)}(k)=\varepsilon_2T_2(k)-\dfrac{1}{\pi}
\int\limits_{0}^{\infty} J(k,k_1)E_1(k_1)dk_1,\cdots.
\eqno{(25)}
$$

We call the determinant of the dispersion matrix the dispersion function:
$$
\lambda(z)\equiv\det \Lambda(z)=k^2\omega(k),
$$
where
$$
\omega(k)=\dfrac{3}{g_1 g_2}
\Big[k^2T^{3,0}_{\gamma+4,6}(k)T^{2,2}_{\gamma+4,2}(k)+
T^{1,1}_{2\gamma+4,2}(k)T^{2,1}_{2\gamma+4,2}(k)\Big].
$$

The inverse  matrix  to the dispersion matrix is
$$
\Lambda^{-1}(k)=\dfrac{D(k)}{\lambda(k)},
$$
where
$$
D(k)=\left [\extrarowheight=18pt  \begin{array}{cc}
\dfrac{k^2}{g_2}T^{2,2}_{\gamma+4,2}(k)
 & -\dfrac{ik}{g_2}T^{1,1}_{2\gamma+4,2}(k) \\
   - \dfrac{3ik}{g_1}T^{2,1}_{2\gamma+4,2}(k) &
   \dfrac{3k^2}{g_1}T^{3,0}_{\gamma+4,6}(k)   \\
  \end{array}
\right].
$$

We consider the construction of series (20).
We designate
$$
E^{(m)}(k)=\left[\extrarowheight=10pt\begin{array}{c}
           E_1^{(m)}(k) \\
           E_2^{(m)}(k) \\
         \end{array}
       \right], \qquad m=0,1,2,\cdots.
$$

From the equation (23) we obtain
$$
E_1^{(0)}(k)=\dfrac{ik}{\omega(k)g_2}
T^{2,2}_{\gamma+4,2}(k)\Big[B^+T^{1,0}_{2\gamma+4,4}(k)+
\varepsilon_0T^{1,0}_{2\gamma+5,3}(k)\Big]-$$$$
+\dfrac{i}{k\omega(k)g_2}T^{1,1}_{2\gamma+4,2}(k)
\Big[B^+T^{0,1}_{3\gamma+2,2}(k)-\varepsilon_0T^{0,1}_{3\gamma+3,1}(k)\Big],
\eqno{(26)}
$$
and
$$
E^{(0)}_2(k)=\dfrac{3}{\omega(k)g_1}T^{2,1}_{2\gamma+4,2}\Big[
B^+T^{1,0}_{2\gamma+4,4}-\varepsilon_0T^{1,0}_{2\gamma+5,3}(k)\Big]+
$$
$$
+\dfrac{3}{\omega(k)g_1}T^{3,0}_{\gamma+4,6}
\Big[-B^+T^{0,1}_{3\gamma+2,2}(k)+
\varepsilon_0T^{0,1}\alpha_{3\gamma+3,1}(k)\Big].
\eqno{(27)}
$$

The quantity $E^{(0)}_2(k)$ exists for all values of $k$ (has no
singularities).
But quantity $E_1^{(0)}(k)$ has a simple pole at
point $k=0$. Eliminating the singularity at zero, we take
$$
\varepsilon_0=B^+\dfrac{T^{0,1}_{3\gamma+2,2}(0)}
{T^{0,1}_{3\gamma+3,1}(0)}.
$$

Here
$$
T^{0,1}_{3\gamma+2,2}(0)=\int\limits_{0}^{1}\int\limits_{0}^{\infty}
\E(C)C^{\gamma+2}\mu^2\,d\mu\,dC=\dfrac{1}{3}\int\limits_{0}^{\infty}
\E(C)C^{\gamma+2}\,dC
$$
and
$$
T^{0,1}_{3\gamma+3,1}(0)=\int\limits_{0}^{1}\int\limits_{0}^{\infty}
\E(C)C^{\gamma+2}\mu^2\,d\mu\,dC=\dfrac{1}{2}\int\limits_{0}^{\infty}
\E(C)C^{\gamma+3}\,dC.
$$

We designate now
$$
g_{\varepsilon,2}(\gamma)=\int\limits_{0}^{\infty}\E(C)C^{\gamma+2}
g(C)dC,
$$
$$
g_{\varepsilon,3}(\gamma)=
\int\limits_{0}^{\infty}\E(C)C^{\gamma+3}
g(C)dC,
$$
$$
\E(C)=\sqrt{w_0^2c^2+(C^2/2)^2}, \qquad g(C)=\dfrac{
e^{\E(C)}}{(e^{\E(C)}-1)^2}.
$$

Now we have
$$
\varepsilon_0=
B^+\dfrac{2g_{\varepsilon,2}(\gamma)}{3\gamma_{\varepsilon,3}(\gamma)}.
$$

The received expression for $\varepsilon_0$ we will substitute in
(25) and (26).
Let's substitute expressions (25) and (26) in (23).
From the received equation
we find $E^{(1)}(k)$. Then we will substitute $E^{(1)}(k)$ in
the equation (24). From the received equation we find
$E^{(2)}(k)$. Continuing this process beyond all bounds,
let's construct all members of the series (20) and (21).

\section{\large 7. Temperature jump and Kapitsa resistance}

We will find the quantity $\varepsilon_0$ in an explicit form.
The quantity $B^+$ is proportional to the heat flux:
$$
\mathbf{Q}=\int f(x,\mathbf{p})
\dfrac{\partial \E(p)}{\partial \mathbf{p}}
\,\E(p)d\Omega_B.
$$

We transform this expression as
$$
\mathbf{Q}=\dfrac{(2s+1)}{(2\pi \hbar)^3}
\int h(x,\mathbf{p})g(p)\alpha(p)\mathbf{p}\E(p)\,d^3p.
$$

Let's pass in this expression to integration by the dimensionless
to momentum. For this purpose we will notice, that
$$
\alpha(p)\mathbf{p}\,\E(p)\,d^3p=
\dfrac{kT_s}{m}(kT_sm)^2 \alpha(C)\mathbf{C}\E(C).
$$

We receive that
$$
\mathbf{Q}=\dfrac{(2s+1)(kT_s)^3m}{(2\pi \hbar)^3}
\int h(x,\mathbf{C})\alpha(C)\mathbf{C}\E(C)g(C)\,d^3C.
$$

We replace here  the function $h(x,\mathbf{C})$ with its Chapman --- Enskog
expansion $h_{as}(x,\mathbf{C})$. As a result, we have
for the $x$--component of the thermal flux
$$
Q_x=\dfrac{(2s+1)(kT_s)^3m}{(2\pi \hbar)^3}
\int\limits_{-1}^{1}
\int\limits_{0}^{\infty}\int\limits_{0}^{2\pi}
\Big[B^+\alpha(C)C\mu-\varepsilon_T\E(C)\Big]\times $$$$ \times
\alpha(C)C\mu \E(C)C^2g(C) d\mu dC
d\chi=
\dfrac{(2s+1)(kT_s)^3m}{(2\pi \hbar)^3}\cdot
\dfrac{4\pi}{3}g_{\alpha\varepsilon}(\gamma)B^+.
$$

Here
$$
g_{\alpha\varepsilon}(\gamma)=\int\limits_{0}^{\infty}
\alpha^2(C)\E(C)C^4g(C)\,dC.
$$

We hence have
$$
B^+=Q_x\dfrac{6\pi^2\hbar^3}
{(2s+1)m(kT_s)^3g_{\alpha\varepsilon}(\gamma)}.
$$

Thus, the quantity $ \varepsilon_0$ is equal:
$$
\varepsilon_0=Q_x\dfrac{6\pi^2\hbar^3}{(2s+1)m(kT_s)^3}\cdot
\dfrac{g_{\varepsilon,2}(\gamma)}{3g_{\varepsilon,3}(\gamma)
g_{\alpha\varepsilon}(\gamma)}.
\eqno{(28)}
$$

Returning to the formula for the temperature jump
$$
\Delta T=RQ_x,
$$
we find from expression (28) that the
Kapitsa resistance in the zeroth approximation is
$$
R=C(\gamma,q)\dfrac{\hbar^3}{(2s+1)k^3T_s^2m}.
$$

Here
$$
C(\gamma,q)=\dfrac{2\pi^2 g_{\varepsilon,2}(\gamma)}
{g_{\varepsilon,3}(\gamma)g_{\alpha\varepsilon}(\gamma)}\cdot
\dfrac{1+q}{1-q}
\eqno{(29)}
$$
is the (dimensionless) coefficient of the
temperature jump.

The plus sign in formula (28) indicates that the wall temperature
is higher than the phonon component temperature.

\begin{figure}[ht]
\begin{center}
\includegraphics[width=0.49\textwidth]{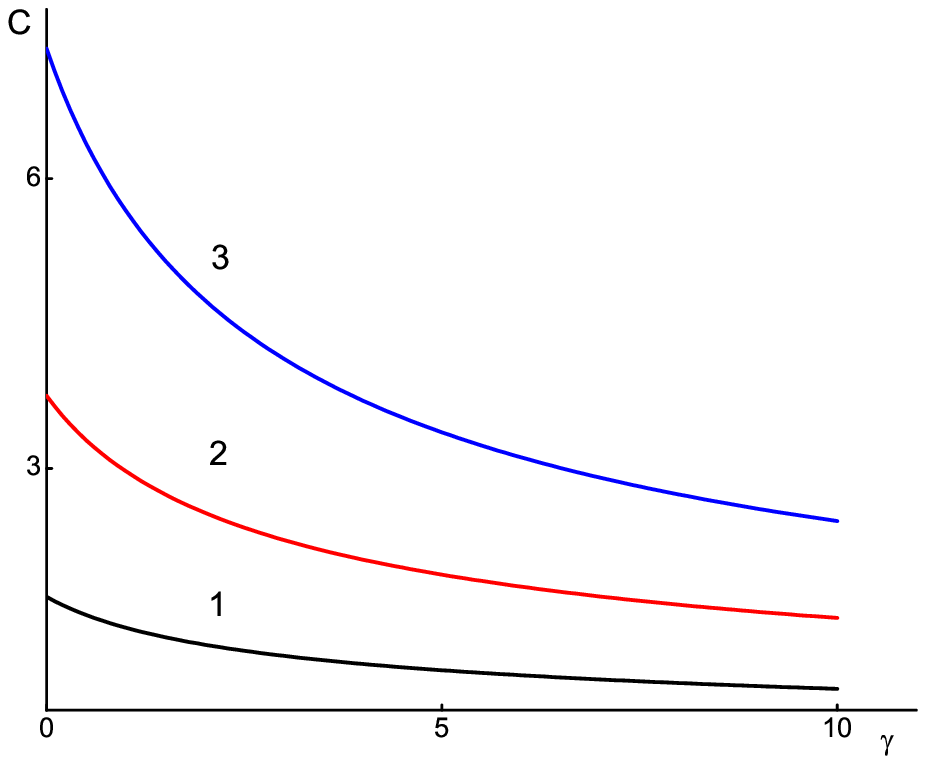}
\hfill
\includegraphics[width=0.49\textwidth]{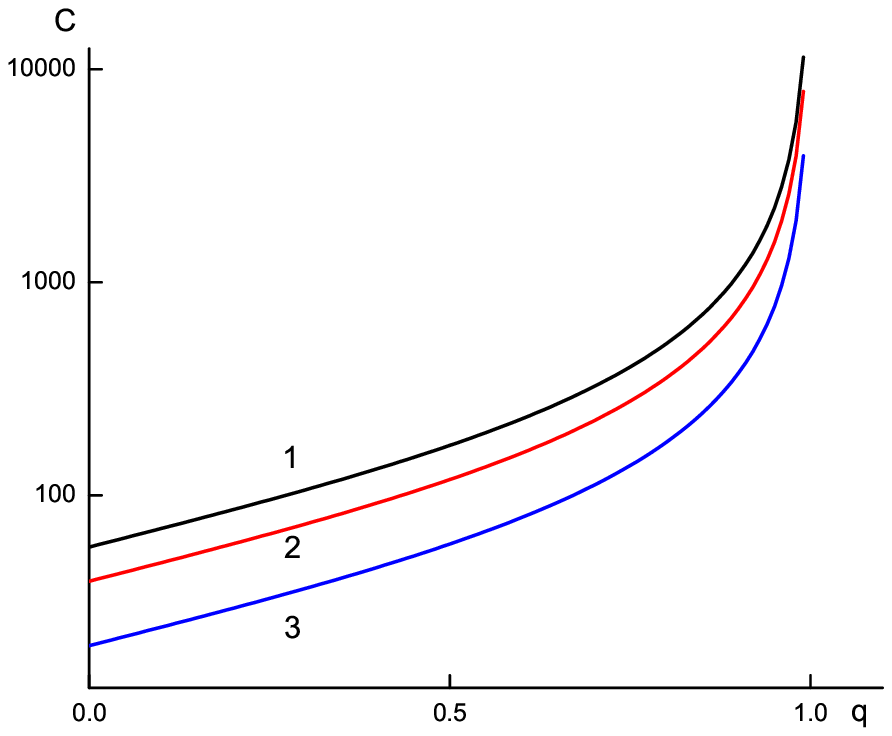}
\\
\parbox[t]{0.49\textwidth}{\normalsize Fig. 1.}
\hfill
\parbox[t]{0.49\textwidth}{\normalsize Fig. 2.}
\end{center}
\end{figure}

The graphs of the behavior of the temperature jump
coefficient as a function of the parameter $\gamma$ and
the specular reflection coefficient $q$ are shown
in Figs. 1 -- 4.

So, on fig. 1 the dependence of temperature jump coefficient
on the parameter $ \gamma $ in case of zero factor
of reflectivity ($q=0$) is presented. Curves of $1,2,3$
correspond to values of dimensionless velocity $w_0=1,2,3$.

\begin{figure}[h]
\begin{center}
\includegraphics[width=0.49\textwidth]{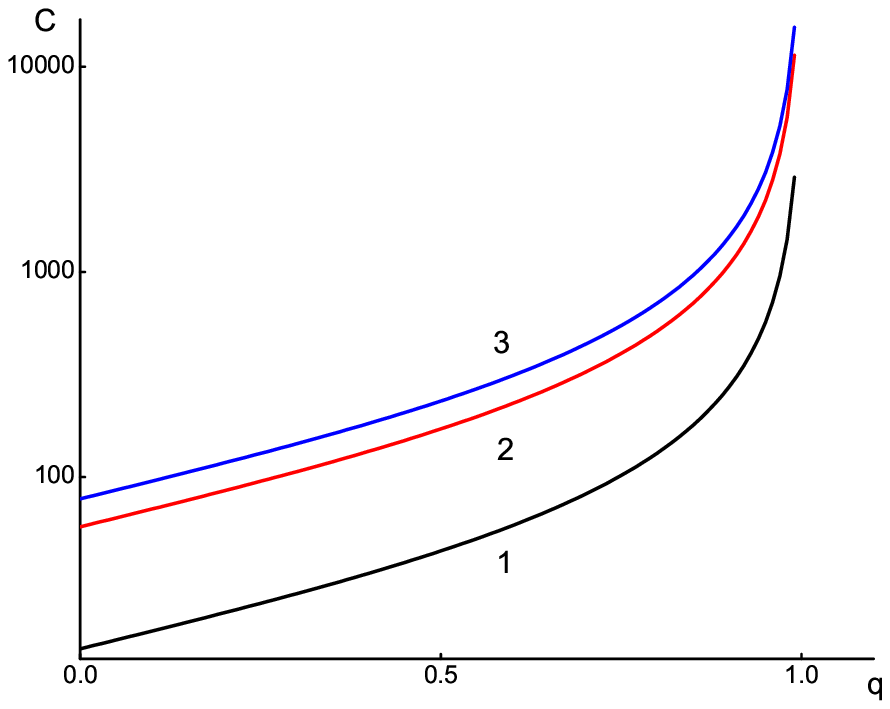}
\hfill
\includegraphics[width=0.49\textwidth]{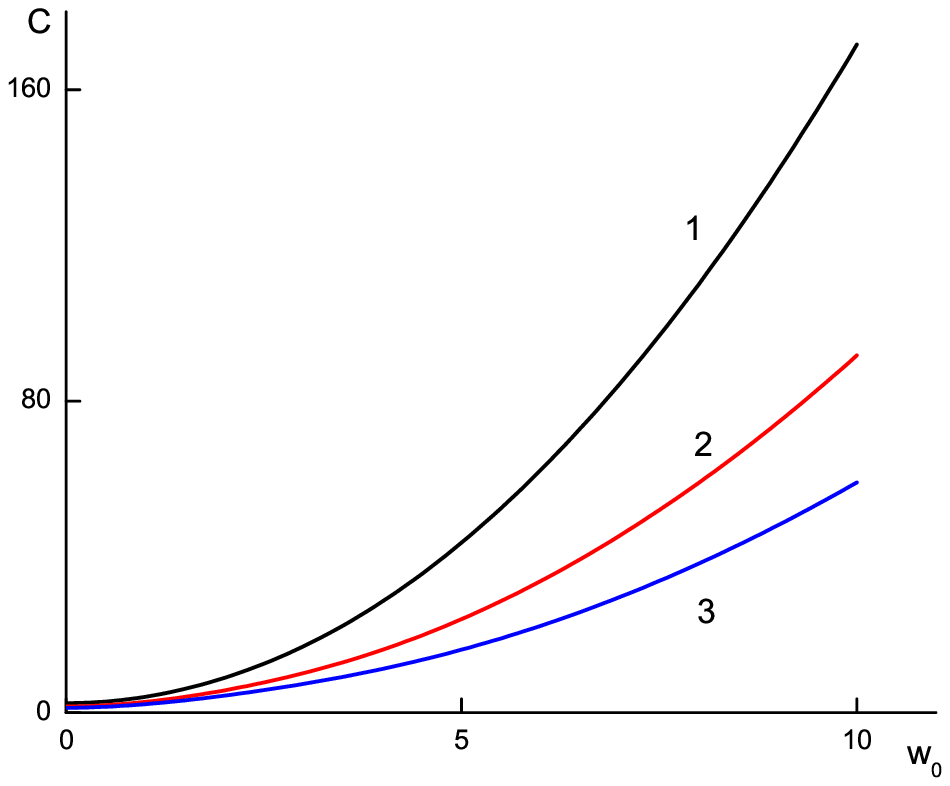}
\\
\parbox[t]{0.49\textwidth}{\normalsize Fig. 3.}
\hfill
\parbox[t]{0.49\textwidth}{\normalsize Fig. 4.}
\end{center}
\end{figure}

On fig. 2 the dependence of this factor on reflectivity factor $q $
in a case, when
dimensionless velocity of the sound $w_0=10$ is presented.
Curves of $1,2,3$ correspond to values of the
parameter of collisions $ \gamma=1,3,10$.

On fig. 3 for the case when $ \gamma=1$ the dependence of the
temperature jump coefficient on reflectivity factor $q$ is given.
Curves of $1,2,3$ correspond to values
dimensionless velocity $w_0=5,10,20$.

On fig. 4 for the case when $q=0.5$ the dependence of the coefficient
on the  dimensionless velocity $w_0$ is shown.
Curves of $1,2,3$ correspond to values of parameter $\gamma=1,5,10$.

From the graphs shown it is seen, that at the fixed values
of the parameters the quantity of the factor $C(\gamma,q)$:

Monotonously decreases with the growth of the paramater $\gamma$,

Monotonously grows with growth of parameter of reflectivity $q $,

Monotonously grows with growth of dimensionless quantity of
velocity of sound $w_0$.

Besides, under convergence of reflectivity coefficient
$q$ to unit the quantity $C(\gamma,q)$
increases unlimitedly, as in this limit the heat exchange between the
wall and gas adjacent to it becomes impossible.

\section{\large 8. Conclusion}

For degenerate quantum bose - gas with frequency of collisions,
depending on momentum of elementary excitations of bose - gas,
the kinetic equation is constructed.
The general case of dependence of energy of elementary excitations
of bose - gas on momentum is considered.
Boundary conditions are assumed to be specular -- diffusive.
The solution of a semi--spatial boundary problem of jump
of temperature on border degenerate bose - gas in the presence
of
Bose --- Einshtein condensate is received. The formula for
finding of jump of temperature and calculation of
Kapitsa  resistance is deduced.
Sufficiently general method of the solution  of
the kinetic equations with specular -- diffusive boundary
conditions,
for the first time offered in \cite {16} in the problem of skin -- effect
is developed.

\end{document}